\documentstyle[preprint,aps,floats,psfig]{revtex}

\begin{document}
 \draft
 \title{Inflation with a constant ratio of scalar and
 tensor perturbation amplitudes}
 \author{C\'esar A. Terrero-Escalante$^{\dagger}$, James E. Lidsey$^*$,
 Alberto A. Garc\'{\i}a$^{\dagger}$}
 \address{$^{\dagger}$Departamento~de~F\'{\i}sica,
 ~Centro de Investigaci\'on y de Estudios Avanzados del IPN,
 ~Apdo.~Postal~14-740,~07000,~M\'exico~D.F.,~M\'exico.\\
 $^*$ Astronomy Unit, School of Mathematical Sciences, Queen Mary,
 University of London, Mile End Road, LONDON, E1 4NS, UK.
 }
 \maketitle

 \begin{abstract}
 The single scalar field
 inflationary models that lead to
 scalar and tensor perturbation spectra with amplitudes
 varying in direct proportion to one another
 are reconstructed by solving the Stewart--Lyth inverse problem to
 next--to--leading order in the slow--roll approximation.
 The potentials asymptote at high
 energies to an exponential form, corresponding to power law
 inflation, but diverge from this model at low energies,
 indicating that power law inflation is a repellor
 in this case. This feature implies that
 a fine--tuning of initial conditions is required if such models
 are to reproduce the observations. The required initial
 conditions might be set through the eternal inflation mechanism.
 If this is the case, it will imply that the spectral indices must be nearly
 constant, making the underlying model observationally indistinguishable
 from power law inflation.

 \end{abstract}


 \section{Introduction}
 \label{sec:Intro}

 A cosmological model with a low density of cold dark matter
 $(\Omega_{\rm CDM} \approx 0.3 )$ and a dominant
 cosmological constant $(\Omega_{\Lambda} \approx 0.7 )$
 is currently favored by recent analyses of
 cosmological observations,
 in particular those of the cosmic microwave background
 (CMB) power spectrum (see for instance, \cite{CMBdata,A_T,TDegen,hudo})
 and high red--shift surveys of type Ia supernovae
 \cite{SIA_O,Perlmutter:1999np}.
 These
 analyses maximize
 the likelihood between
 the observed power spectrum and that
 of a given theoretical model by
 fitting values for various model parameters.
 This set of parameters
 includes quantities that characterize the initial conditions for
 the evolution of the density (scalar)
 and gravitational wave (tensor) perturbations.
 If these initial conditions are
 described by adiabatic, gaussian
 fluctuations,  the
 observations strongly constrain the present--day universe to
 be very nearly spatially flat
 \cite{CMBdata,A_T,TDegen}.

 The simplest casual mechanism for producing a spatially flat universe,
 where large--scale structure
 originates from a primordial spectrum of adiabatic density
 perturbations,
 is given by the single field
 inflationary model
 \cite{inflation}. The early universe undergoes a
 rapid,
 accelerated expansion that is driven by the self--interaction
 potential energy of a
 real scalar field -- {\em
 the
 inflaton}.
 We
 consider inflation driven by a single inflaton field
 throughout this paper.
 The
 field undergoes quantum fluctuations
 as it slowly rolls down its potential and these fluctuations  generate
 a perturbation in the spatial curvature
 \cite{starwave,perturbations}. A primordial
 spectrum of tensor perturbations
 is also generated from the scalar field fluctuations
 \cite{starwave,abbott}.

 Until recently, the majority
 of CMB data analyses have neglected
 the possible effects of the
 primordial gravitational wave spectrum \cite{CMBdata}.
 (For a recent review, see, e.g., Ref. \cite{hudo}).
 In the last few years, however,
 there has been a growing recognition that
 the role of the tensor perturbations
 deserves more attention when determining the
 best--fit values of the
 cosmological parameters \cite{A_T,TDegen,TDegen1}.
 It has been found that if the tensor
 modes are included in the analysis, an extra degeneracy
 arises
 when fitting cosmological models to the
 CMB data \cite{TDegen,TDegen1}.

 Further motivation for including the tensor modes in the analysis
 arises from the possibility that
 a
 measurement of the
 CMB polarization may indirectly determine
 the gravitational wave contribution to the power spectrum \cite{TDegen1,KK}.
 Such a contribution can be parametrized in terms of the quantity
 \begin{equation}
 \label{eq:r}
 r \equiv \alpha\frac{A_T^2}{A_S^2}\, ,
 \end{equation}
 representing the relative amplitudes
 of the tensor $(A_T)$ and scalar $(A_S)$ perturbations,
 where the constant, $\alpha$, depends on the
 particular normalization of the spectral amplitudes that is chosen.

 Given the current accuracy of the observations,
 it is only possible at present to place
 a constant upper
 bound on the allowed range of
 $r$ \cite{A_T,TDegen}.
 Looking to the future,
 the most optimistic expectation
 we have for constraining
 $r$ is that the observations
 would favor a constant central value \cite{TDegen1,KK}.
 It
 is unlikely that a direct dependence of $r$ on scale
 would ever be measurable.

 Motivated by these future expectations, therefore, it is
 important to determine the main features
 of the single field inflationary models that yield perturbation
 spectra resulting in a tensor--to--scalar ratio, $r$,
 that is precisely constant,
 at least over the range of scales accessible to observations.
 This is the purpose of the
 present paper. Of particular interest is the functional form of
 the corresponding inflaton potentials.
 Such a study is also relevant to the question of whether
 the inflaton potential can be
 reconstructed directly from observations and, consequently,
 provides important information about physics at very high energy
 scales \cite{Lea}.

 It follows immediately from the definition (\ref{eq:r})
 that imposing the condition $r={\rm constant}$
 is equivalent to searching for models where the scale dependences
 of the two spectra are identical.
 In general, these scale dependences
 are parametrized in terms of
 the scalar ($n_S$) and tensor ($n_T$) spectral indices, respectively.
 These are defined in
 terms of the logarithmic
 derivatives of the corresponding normalized amplitudes:
 \begin{eqnarray}
 \label{eq:nSDef}
 \Delta&\equiv&\frac{n_S-1}2 \equiv \frac{d\ln A_S}{d\ln k} \, ,\\
 \label{eq:nTDef}
 \delta&\equiv&\frac{n_T}2 \equiv \frac{d\ln A_T}{d\ln k} \, ,
 \end{eqnarray}
 where $k=aH$ is the comoving wavenumber when the mode
 first crosses the Hubble radius, $d_H\equiv H^{-1}$, during inflation.

 We are therefore interested in the models that produce
 \begin{equation}
 \label{interest}
 \Delta (k) = \delta (k) .
 \end{equation}
 Since exact expressions relating
 the perturbation spectra to the inflaton potential
 are presently unknown, the standard approach is to expand the
 power spectra
 in terms
 of a set of parameters that
 describe the inflationary dynamics through the slow evolution
 of the Hubble radius or, in the case
 of single scalar field models,
 in terms of the
 slow rolling
 of the inflaton field
 along a nearly flat potential (see
 Refs.~\cite{inflation,Lea,HFFampl} and references therein). To leading order
 in
 these expansions, the inflationary models that have precisely
 constant ratio, $r$, are precisely the power law inflationary
 models driven by an exponential potential \cite{PLinfl}.
 For power law inflation the spectral indices are constant
 and equal to each other.
 However, as we discuss in the next section,
 to next-to-leading order,
 the indices can vary with scale whilst
 being equal to each other at each value of $k$.
 Specifically,
 we solve the next--to--leading order
 Stewart-Lyth inverse problem (SLIP) \cite{SLIP}
 under the condition that the spectral indices
 are equal at any scale. We find
 that for
 this case, power law behavior is a repellor rather than an attractor of
 the
 inflationary dynamics, in contrast to the
 lowest--order analysis of  Hoffman and
 Turner \cite{HT}. This difference arises
 because different approximations
 in the slow--roll analysis are considered.
 Nevertheless, if the underlying inflaton potentials
 derived at next--to--leading order
 are to
 produce successful inflation under
 the condition that $r={\rm constant}$ $(\Delta =\delta )$,
 a strong fine--tuning of the initial
 value
 for the inflaton field
 is required.
 The eternal inflation mechanism \cite{eternal} provides a
 natural way of obtaining the required initial values,
 although
 in this case, the scale dependence of the spectral indices
 is strongly suppressed and, consequently, the difference
 between these models and
 the power law model is effectively erased.

 The paper is organized as follows. We present the
 equations that determine the perturbation spectra in Section II.
 In Section III, we proceed to solve these equations
 and reconstruct the corresponding inflationary potentials
 in parametric form. We conclude with a discussion in Section IV.

 \section{Perturbation Spectra}
 \label{sec:pre}

 When determining the cosmological
 parameters \cite{CMBdata,A_T,TDegen},
 the primordial perturbation spectra are often expanded as
 \begin{eqnarray}
 \label{eq:SExp}
 \ln A_S^2(k)&=&\ln A_S^2(k_*) + \Delta(k_*)\ln\frac{k}{k_*}
 + \frac12\frac{d\Delta(k)}{d\ln k}\vert_{k=k_*}\ln^2\frac{k}{k_*} + \cdots
 \, ,\\
 \label{eq:TExp}
 \ln A_T^2(k)&=&\ln A_T^2(k_*) + \delta(k_*)\ln\frac{k}{k_*}
 + \frac12\frac{d\delta(k)}{d\ln k}\vert_{k=k_*}\ln^2\frac{k}{k_*} + \cdots
 \, ,
 \end{eqnarray}
 where $k_*$ is a pivotal scale.
 The order where these expansions are truncated
 is determined by the precision of the CMB observations.
 Only the first two terms in each expansion are usually taken into account
 and the `running'  of the indices,
 $dn_i /d\ln k$,
 is neglected. Even so,
 there are still
 four inflationary parameters that need
 to be fitted to the data: $A_S(k_*)$, $A_T(k_*)$,
 $\Delta(k_*)$ and $\delta(k_*)$.
 On the
 other hand, the total number of
 cosmological parameters in the analysis
 can be very large \cite{hudo},
 and in view of the level of accuracy expected from
 forthcoming observations in the near future,
 it is possible that
 higher--order
 terms in the expansions (\ref{eq:SExp}) and (\ref{eq:TExp})
 may have to be included as well.
 The large  number of parameters that need to be considered
 allows
 degeneracies to arise when fitting  a given theoretical model
 to the observational data.

 One way of reducing the number of degrees of freedom is
 to fit the ratio, $r$, instead
 of the tensor quantities directly.
 The relevant tensor modes can then be deduced
 from
 the
 definition of $r$,   Eq.~(\ref{eq:r}),
 and the next-to-leading order consistency relation
 for the class of single field models
 \cite{Lea}:
 \begin{equation}
 n_{\rm T}  = -2 \left[
 \frac{r}{\alpha} - \left( \frac{r}{\alpha}
 \right)^2 - \left( n_{\rm S} - 1 \right)
 \left( \frac{r}{\alpha} \right) \right] .
 \end{equation}
 The bounds \cite{TDegen,TDegen1,KK}
 in the observed precision of the tensor contribution to
 the CMB power spectrum are such that
 even after a best--fit value for
 $r(k_*)$ has been deduced,
 the simplest approach
 to adopt when constraining the parameter space
 would be to assume that the value $r(k_*)$
 holds for all scales,
 i.e, $r(k)=r(k_*)={\rm constant}$.
 As we have already seen,
 this corresponds formally to a model with
 $\Delta(k)=\delta(k)$. Consequently,
 this procedure provides further
 motivation for analyzing the case where $r$ is precisely
 constant, because this allows us in principle to gauge to what
 extent the accuracy of
 expansions (\ref{eq:SExp}) and (\ref{eq:TExp})
 is sensitive to the
 value of $k_*$.

 In single--field
 inflation,
 the indices,
 $\Delta (k) $ and $\delta (k)$, satisfy the SLIP
 equations \cite{SLIP}:
 \begin{eqnarray}
 \label{eq:MSch1}
 2C\epsilon_1 \hat{\hat{\epsilon_1}}-(2C+3)\epsilon_1 \hat{\epsilon_1}
 -\hat{\epsilon_1}
 +\epsilon_1^{2}+\epsilon_1 +\Delta &=&0\,,  \\
 \label{eq:MSch2}
 2(C+1)\epsilon_1\hat{\epsilon_1}-\epsilon_1^{2}-\epsilon_1 -\delta
 &=&0\,,
 \end{eqnarray}
 to `next--to--leading order'
 in the slow--roll approximation,
 where $C=-0.7296$ and a circumflex accent denotes
 differentiation with respect to
 the variable $\tau$, defined such that
 $d\tau\equiv d\ln H^2$. The first `horizon flow function'
 is defined as \cite{HFFampl,Lea}
 \begin{equation}
 \label{eq:e1}
 \epsilon_1 \equiv {{\rm d} \ln d_{\rm H}\over {\rm d} N} =
 \frac{3T}{T+V}
 \end{equation}
 where $T\equiv \dot{\phi}^2 /2$
 represents the kinetic energy of the inflaton field,
 and
 $N \equiv \ln (a/a_{\rm i})$ is the number of {\em e}--foldings
 of inflationary expansion since some initial time,
 $t_{\rm i}$. (Note that the
 number of {\em e}--foldings is usually counted backwards in time.
 Here, we count it forward,
 i.e., $N(t_{\rm i})=0$).  In general,
 Eq. (\ref{eq:e1})
 measures the logarithmic change of the Hubble distance per {\em e}--folding,
 or equivalently,
 the contribution of the inflaton field's
 kinetic energy relative to its total energy density.
 Inflation proceeds for $\epsilon_1 < 1$
 ($\ddot{a} > 0$) and
 the weak energy condition for a spatially
 flat universe
 is satisfied for
 $\epsilon_1 >0$.
 By combining  Eqs.~(\ref{eq:MSch1})
 and (\ref{eq:MSch2})
 we deduce that
 \begin{equation}
 \label{eq:Meq}
 \delta-\Delta=2C\epsilon_1\hat{\hat{\epsilon_1}}-(\epsilon_1+1)
 \hat{\epsilon_1} \, .
 \end{equation}

 In the next Section, we proceed to find solutions to Eq. (\ref{eq:Meq}).

 \section{The model}
 \label{sec:model}

\begin{figure}[ht]
 \centerline{\psfig{file=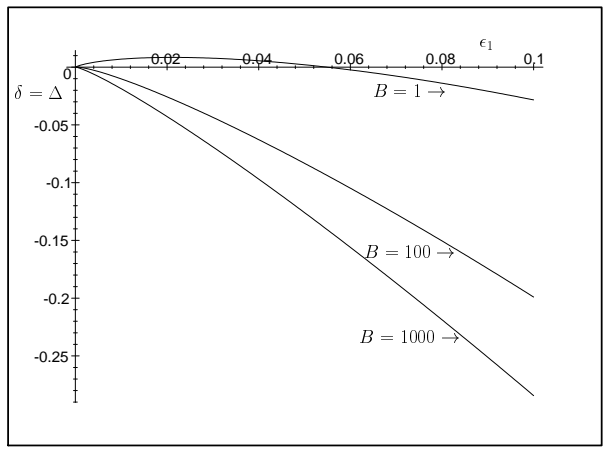,width=8.5cm}}
 \caption{The variation
 of the scalar and tensor spectral indices,
 $\delta=\Delta$, as a function of the first horizon
 flow function $\epsilon_1$ for different values of the constant $B$.}
 \label{fig:GPLdDe}
 \end{figure}

 When determining the single field models that lead to
 $r={\rm constant}$ we are interested in solving Eq. (\ref{eq:Meq})
 when Eq. (\ref{interest}) is imposed.
 Firstly, we remark that
 one solution to Eq.~(\ref{eq:Meq})
 with this condition is that
 $\epsilon_1$, $\Delta$ and  $\delta$ are all constants,
 corresponding to the power law inflationary model
 \cite{PLinfl}. Moreover, power law inflation is
 the {\em unique}
 solution if second--order terms in Eq.~(\ref{eq:Meq}) are neglected.
 More generally,
 a first integration of Eq.~(\ref{eq:Meq}) yields
 \begin{equation}
 \label{eq:hatEps}
 \hat{\epsilon_1}= \frac{1}{2C}
 \left[\epsilon_1+\ln\left(B\epsilon_1\right)\right]
 \, ,
 \end{equation}
 when
 Eq. (\ref{interest}) is satisfied,
 where $B>0$ is an integration constant. Substituting this result into
 Eq.~(\ref{eq:MSch2}) we find that
 \begin{equation}
 \label{eq:Delta_Eps}
 \Delta(\epsilon_1)=\delta(\epsilon_1)
 =
 \frac{C+1}{C}\epsilon_1\left[\epsilon_1+\ln\left(B\epsilon_1\right)\right]
 - \epsilon_1^2 - \epsilon_1
 \, .
 \end{equation}
 Eq. (\ref{eq:Delta_Eps}) is plotted in  Fig. \ref{fig:GPLdDe}
 for different values of $B$.

 Now, it has been shown in Ref.
 \cite{ConstNt} that to next--to--leading order,
 the constraint
 $\delta(\epsilon_1)\leq 0$ must be satisfied for any value of
 $\epsilon_1$ during inflation.
 However,
 as indicated in
 Fig.~\ref{fig:GPLdDe},
 Eq. (\ref{eq:Delta_Eps}) is
 a generalized inverted parabola and always
 has a
 positive
 maximum. This implies that $\delta (\epsilon_1 )>0$
 for some range of $\epsilon_1$.
 Indeed, the positive maximum
 is located in the interval $\epsilon_1\in(0,1/B)$.
 This would
 seem to indicate that the condition for
 the spectral indices to be equal (but not constant)
 might not be self--consistent at this level of
 the slow--roll approximation. However, the point is that
 $\delta<0$ is necessary only during inflation and,
 in effect,
 this condition further constrains the
 allowed range of values that $\epsilon_1$
 may take for the analysis to be self--consistent. Indeed,
 we may specify the parameter, $B$, in such a way that
 the values $\delta(\epsilon_{1(max)})$ and $\epsilon_{1(max)}$ are negligible,
 where
 \begin{equation}
 \label{eq:EpsMax}
 \epsilon_{1(max)} = \frac{C+1}2
 {\mathcal
 L}_W\left[\frac{2\exp\left(-\frac{1}{C+1}\right)}{B(C+1)}\right]
 \end{equation}
 is the value of $\epsilon_1$ at the maximum value of
 $\delta(\epsilon_1)$ and ${\mathcal L}_W$ is the Lambert $W$ function
 \cite{LW}.
 As we shall see later, $1/B$ determines the range of power law
 behavior of
 $\Delta$. Thus, taking
 into account the range of $n_S$ consistent
 with analyses of CMB observations \cite{CMBdata,A_T,TDegen},
 we can choose $B=100$.
 In this way, $\epsilon_{1(max)}\approx 0.00025$
 can be identified as the
 lower
 critical value of $\epsilon_1$ where
 the precision in
 our calculations is consistent to
 next-to-leading order in the slow-roll approximation,
 with $\delta(\epsilon_{1(max)})\approx 0.00009$
 then being
 consistently negligible.

 Before proceeding, let us first analyze the
 quantity
 \begin{equation}
 \label{eq:den}
 Q \equiv \frac{1}{\epsilon_1^2+\epsilon_1+\delta}
 \, ,
 \end{equation}
 since this term plays an
 important role
 in the forthcoming calculations. Substituting $\delta$, as given by
 Eq. (\ref{eq:Delta_Eps}), into Eq.~(\ref{eq:den})
 implies that
 \begin{equation}
 \label{eq:den1}
 Q = \left(\frac{C}{C+1}\right)
 \frac{1}{\epsilon_1\left[\epsilon_1+\ln\left(B\epsilon_1\right)\right]}
 \, .
 \end{equation}
 In order to obtain analytical results, it is necessary
 to approximate the denominator of Eq. (\ref{eq:den1})
 in the different limits where
 $\epsilon_1 \gg \left| \ln \left(
 B \epsilon_1 \right) \right|$ and
 $\epsilon_1 \ll \left| \ln \left( B \epsilon_1
 \right) \right|$, respectively. In the former case,
 Eq. (\ref{eq:den1}) can be
 be expanded as the series
 \begin{equation}
 \label{eq:app1den}
 \left(\frac{C}{C+1}\right)
 \frac{1}{\epsilon_1^2}
 \left\{1-\frac{\ln\left(B\epsilon_1\right)}{\epsilon_1}
 +\left[\frac{\ln\left(B\epsilon_1\right)}{\epsilon_1}\right]^2
 - \cdots \right\}
 \end{equation}
 if $\epsilon_1$ is sufficiently different from zero.
 However, since $\epsilon_1$ is typically small during inflation,
 $B$ must be appropriately tuned
 if this expansion is to be valid. Consequently,
 the
 approximation is only
 appropriate over a very narrow range of $\epsilon_1$.
 Indeed, a numerical investigation confirms that the
 relevant range
 of $\epsilon_1$ is negligible for realistic values of $B$
 and so we regard it as unphysical.
 For
 example, if $B=100$, this approximation is valid only for
 $\epsilon_1\in(0.009999,0.010001)$.

 The
 more interesting case is the limit,
 $\epsilon_1 \ll \left|\ln \left( B\epsilon_1
 \right) \right|$, where
 the following approximation for
 expression (\ref{eq:den1}) applies:
 \begin{equation}
 \label{eq:app2den}
 Q = \left(\frac{C}{C+1}\right)
 \frac{1}{\epsilon_1\ln\left(B\epsilon_1\right)}
 \left\{1-\frac{\epsilon_1}{\ln\left(B\epsilon_1\right)}
 +\left[\frac{\epsilon_1}{\ln\left(B\epsilon_1\right)}\right]^2
 - \cdots \right\}
 \, .
 \end{equation}
 Firstly, if
 $\epsilon_1<1/B$, this implies that $\epsilon_1$ is very small and the
 absolute value of $\ln(B\epsilon_1)$ will be very large\footnote{We
 are assuming implicitly that $B \gg 1$ in this discussion.}.
 Secondly,
 if $\epsilon_1>1/B$, then $\epsilon_1\in(1/B,1]$ implies that
 $\ln(B\epsilon_1)>\epsilon_1$ if $B$ is sufficiently
 large. Note that the value of $B$ specifies the central value of
 $\epsilon_1$ during inflation and
 this is expected to be less than $0.1$.
 With this in mind, a numerical test
 implies that the expansion is valid if $\epsilon_1
 > 0.112$ for $B=10$
 and $\epsilon_1 > 0.0101$ if $B=100$.
 Furthermore, one may verify that this approximation
 is suitable even for $\epsilon_1$
 close to $1/B$. For example, when $B=100$,
 the reliable intervals of
 $\epsilon_1$ are $(0,0.0095]$ and $[0.0101,1]$, respectively.

 We now employ this approximation to find
 semi--analytical expressions  for the spectral
 indices.
 The wavenumber, $k =aH$, at horizon crossing
 is evaluated as a function of
 $\epsilon_1$ by solving the first--order
 differential equation \cite{SLIPk}
 \begin{equation}
 \label{eq:ediffk}
 (C+1)(\epsilon_1-1)\tilde{\epsilon}_1-\epsilon_1^2-\epsilon_1-\delta=0,
 \end{equation}
 where $\tilde{\epsilon}_1 \equiv d\epsilon_1/d\ln k$ . Using
 approximation
 (\ref{eq:app2den}) and integrating implies that
 \begin{eqnarray}
 \label{eq:lnk}
 \ln \frac{k}{k_0}
 &=&C\left\{ \left[\frac1{2\ln^2(B\epsilon_1)}
 + \frac2{\ln(B\epsilon_1)}\right]\epsilon_1^2
 - \frac{\epsilon_1}{\ln(B\epsilon_1)}
 + \left[-\frac2B{\mathcal E}_i(1,\mp \ln(B\epsilon_1)) \right. \right.
 \nonumber \\
 &+& \left. \left.
 \frac4{B^2}{\mathcal E}_i(1,\mp 2\ln(B\epsilon_1))
 - \frac9{2B^3}{\mathcal E}_i(1,\mp 3\ln(B\epsilon_1))
 - \ln\left|\ln(B\epsilon_1)\right|\right]\right\} \, ,
 \end{eqnarray}
 where ${\mathcal E}_i(n,x)$ represents the exponential integral
 \cite{Ei},
 a $\mp$
 corresponds to $\epsilon_1<1/B$ and $\epsilon_1>1/B$,  respectively,
 and a sub--dominant term has been consistently neglected.
 The integration constant is denoted by\footnote{Integration
 constants are labeled by a subscript $0$
 in the corresponding variable in what follows.} $k_0$.

 \begin{figure}[ht]
 \centerline{\psfig{file=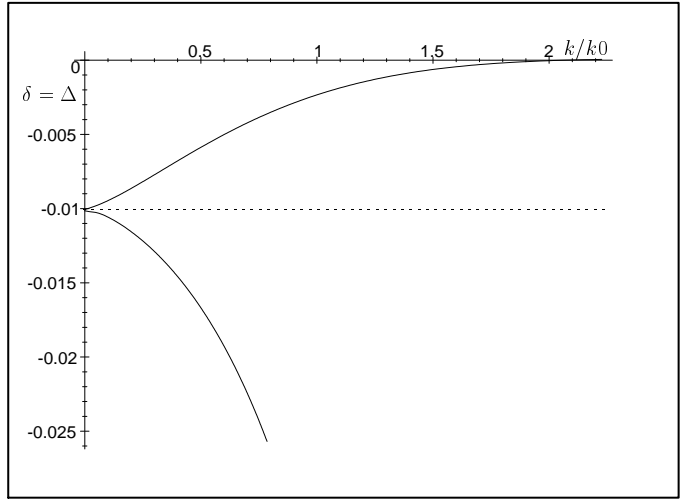,width=9.5cm}}
 \caption{The variation
 of the scalar and tensor spectral indices,
 $\Delta=\delta$, as a function of the
 Hubble radius crossing wavenumber, $k$. The upper branch corresponds to
 $\epsilon_1<1/B$
 and the lower branch to $\epsilon_1>1/B$. The power law solution,
 corresponding
 to $\epsilon_1=1/B$, is represented by the dashed line.}
 \label{fig:GPLdDk}
 \end{figure}

 The dependence
 of the spectral indices,
 $\delta(k)=\Delta(k)$, on comoving wavenumber is
 presented\footnote{Hereafter the plots in this paper
 are consistently drawn using $\epsilon_{1(max)}$
 as the lower value for $\epsilon_1$.} in Fig.~\ref{fig:GPLdDk}.
 It is observed that when the inflationary dynamics
 results in scalar and tensor perturbation spectra
 that satisfy $r={\rm constant}$,
 the power law
 inflationary model, as represented by the dashed line,
 is a {\em repellor}
 rather than an attractor
 to next--to--leading order.

\begin{figure}[ht]
 \centerline{\psfig{file=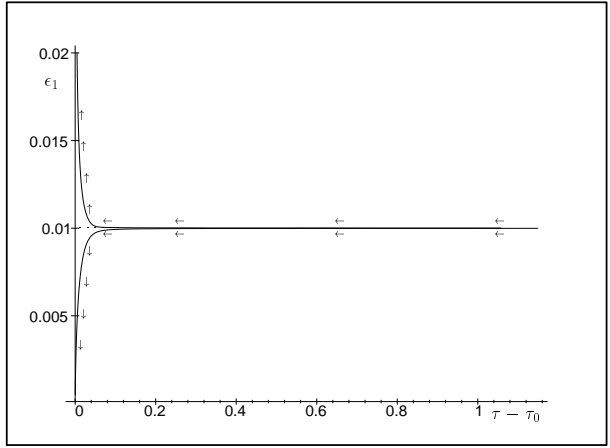,width=8.5cm}}
 \caption{The variation of  $\epsilon_1$ as a function of $\tau - \tau_0
 \equiv \ln H^2$. The flow of time along
 each branch is represented by the small arrows.}
 \label{fig:GPLet1}
 \end{figure}

 It is necessary to analyze the
 dynamics of the parameter, $\epsilon_1$,
 in order to understand this behavior
 more fully. The solution to Eq.~(\ref{eq:MSch2})
 is given by
 \begin{eqnarray}
 \label{eq:C1te}
 \tau&=& \exp\left\{2C\left[
 \frac{\epsilon_1^2}{\ln(B\epsilon_1)}
 -\frac1B{\mathcal E}_i(1,\mp \ln(B\epsilon_1)) \right. \right.
 \nonumber \\
 &+&\left. \left.
 \frac2{B^2}{\mathcal E}_i(1,\mp 2\ln(B\epsilon_1))
 - \frac9{2B^3}{\mathcal E}_i(1,\mp 3\ln(B\epsilon_1))
 \right]
 \right\} + \tau_0
 \end{eqnarray}
 when the approximation (\ref{eq:app2den}) is
 valid.
 Eq. (\ref{eq:C1te}) is plotted in Fig.
 \ref{fig:GPLet1}.

 It follows from this
 figure
 that the two  branches in Fig.~\ref{fig:GPLdDk}
 correspond to initial values of $\epsilon_1$
 that are greater than or less than
 $1/B$, respectively.
 Since $d\tau/dt<0$, the
 solutions diverge from
 the solution given by $\epsilon_1=1/B$ as
 cosmic time, $t$, increases
 and this is
 illustrated in
 Fig.~\ref{fig:GPLet1} by the direction of the
 small arrows near
 to each of the branches.
 In order to emphasize the (inverse) asymptotic
 behavior in the neighborhood of
 the unstable fixed point $\epsilon_1=1/B$,
 the initial values for $\epsilon_1$
 for this figure were
 chosen to be extremely close to $1/B$. However, it can be
 seen that any small deviation of the initial value from $1/B$
 is
 exponentially amplified and
 consequently, the time
 interval that the solution spends inside any given
 neighborhood of $\epsilon_1=1/B$ is exponentially
 suppressed. This implies that,
 unless the initial value of $\epsilon_1$ is extremely close to $1/B$,
 this parameter will immediately
 move out of the $[0,1)$ interval and consequently,
 this leads to a very rapid and
 undesirable end to the
 inflationary period.
 Observe that, in these cases, $\tau_0$
 determines the energy scale, $H_0$, at the end of the inflationary era.
 Returning to Fig.~\ref{fig:GPLdDk}, such a fine tuning of the initial
 conditions for $\epsilon_1$ is translated into
 the requirement that
 a sufficiently large value of the integration
 constant, $k_0$,  must be chosen in order to have
 near power law behavior in the appropriate range of scales.

\begin{figure}[ht]
 \centerline{\psfig{file=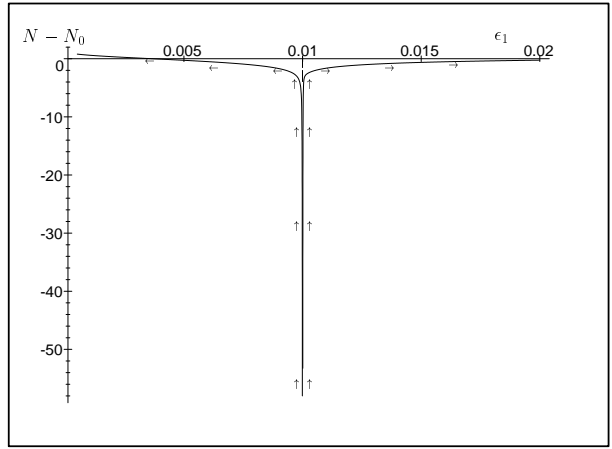,width=8.5cm}}
 \caption{Illustrating the variation
 of the number of {\em e}--foldings of inflationary expansion,
 $N$, as a function of $\epsilon_1$. Time flow along
 each branch is represented by the small arrows.}
 \label{fig:GPLet2}
 \end{figure}

 Further support for this
 conclusion regarding
 the necessary fine-tunning of the initial conditions
 is given by considering the
 variation of the {\em e}--foldings,
 $N$, with respect to $\epsilon_1$. The lower limit,
 $N>60$, must be satisfied if
 inflation is to resolve the horizon and flatness
 problems
 \cite{inflation} and the dependence of the
 number of {\em e}--foldings on $\epsilon_1$
 is determined from the
 differential equation,
 $\hat{N}=-1/2\epsilon_1$ \cite{generic} or,
 equivalently, by evaluating the integral
 \begin{equation}
 \label{eq:DNintde}
 N=-(C+1)\int \frac{d\epsilon_1}{\epsilon_1^2+\epsilon_1+\delta} + N_0
 \, .
 \end{equation}
 In our case, the integration of Eq.~(\ref{eq:DNintde}) yields
 \begin{eqnarray}
 \label{eq:DNe}
 N=&-&C\left\{\left[
 -\frac12\frac1{\ln^2(B\epsilon_1)}
 -\frac1{\ln(B\epsilon_1)}\right]\epsilon_1^2
 + \frac{\epsilon_1}{\ln(B\epsilon_1)}
 +\ln\left|\ln(B\epsilon_1)\right| \right.\nonumber \\
 &+&\left.
 \frac1B{\mathcal E}_i(1,\mp \ln(B\epsilon_1))
 - \frac2{B^2}{\mathcal E}_i(1,\mp 2\ln(B\epsilon_1))
 \right\} + N_0
 \, ,
 \end{eqnarray}
 and the graph corresponding to Eq.~(\ref{eq:DNe})
 is shown in Fig.~\ref{fig:GPLet2}.
 As expected, the relative expansion rate of the universe is
 larger as
 the initial value of
 $\epsilon_1$ becomes closer to
 $1/B$.


\begin{figure}[ht]
 \centerline{\psfig{file=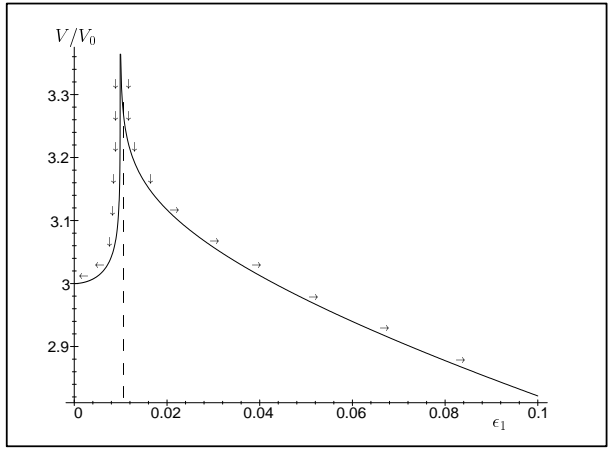,width=8.5cm}}
 \caption{The inflaton potential as function of $\epsilon_1$.}
 \label{fig:GPLVepe1}
 \end{figure}

 The scalar field and its self--interaction potential
 are given in terms of
 $\epsilon_1$ by \cite{ConstNt},
 \begin{eqnarray}
 \label{eq:PotentialE}
 V(\epsilon_1)&=& \frac{1}{\kappa}\left(3-\epsilon_1\right)
 \exp\left[\tau(\epsilon_1)\right]
 \, ,
 \\
 \label{eq:Phi}
 \phi(\epsilon_1)&=& -\frac{2(C+1)}{\sqrt{2\kappa}}
 \int\frac{\sqrt{\epsilon_1}d\epsilon_1}{\epsilon_1^2+\epsilon_1+\delta}
 + \phi_0\, .
 \end{eqnarray}
 and the potential  is shown in
 Fig.~\ref{fig:GPLVepe1}, where $V_0=\kappa^{-1}\exp(\tau_0)$,
 $\kappa = 8\pi/m_{\rm Pl}^2$ is the
 Einstein constant and $m_{\rm Pl}$ is the Planck mass.
 The right--hand branch of the potential has a minimum at
 $\epsilon_1^{(upper)}\approx 0.79$, not shown in the figure in order to allow
 for the observation of details in a more realistic range of values for
 $\epsilon_1$.

\begin{figure}[ht]
 \centerline{\psfig{file=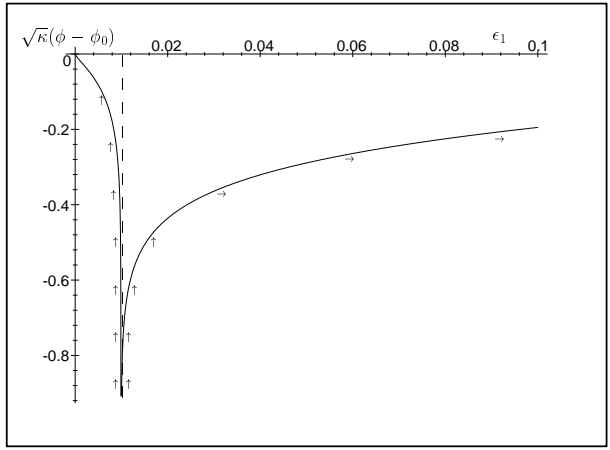,width=8.5cm}}
 \caption{The inflaton field as function of $\epsilon_1$.}
 \label{fig:GPLVepe2}
 \end{figure}

 Unfortunately, even under the approximation (\ref{eq:app2den}),
 it is not possible
 to analytically integrate Eq.~(\ref{eq:Phi}) for the inflaton
 field.
 Nevertheless, this integral can be performed numerically and the result
 is
 presented in Fig.~\ref{fig:GPLVepe2}.

\begin{figure}[ht]
 \centerline{\psfig{file=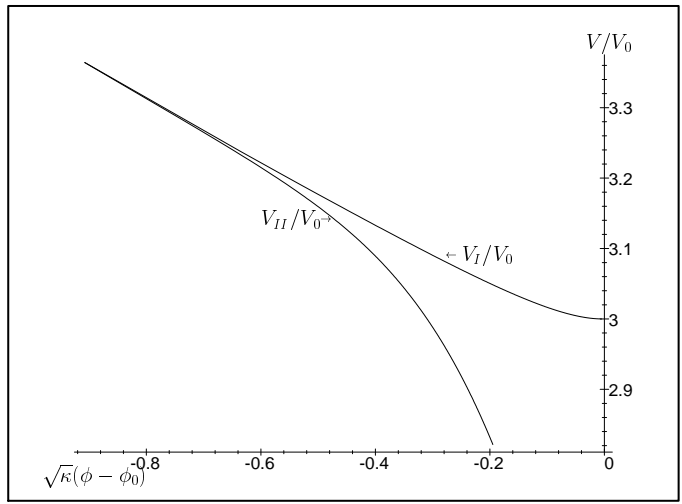,width=9.5cm}}
 \caption{The potential as function of the inflaton field.
 $V_I$ denotes
 the branch where $\epsilon_1\in (0,1/B)$, while $V_{II}$ stands for
 $\epsilon_1\in (1/B, \epsilon_1^{(upper)}]$. Larger values of the potential
 are associated with $\epsilon_1$ being progressively nearer to the
 critical value $1/B$. In this limit, the model approaches the
 power law inflation behavior.}
 \label{fig:GPLVph}
 \end{figure}

 In Eq. (\ref{eq:Phi}), a sign for the square root of $\epsilon_1$
 must be chosen and this specifies the signs
 of $\dot{\phi}$ and $dV/d\phi$.
 We have
 assumed that $\dot{\phi} >0$ and $dV/d\phi <0$ and
 so for consistency, one can only consider those intervals
 of $\epsilon_1$ where this is valid.
 With the above results it is possible to
 check that the necessary criteria \cite{ConstNt},
 \begin{equation}
 \left\{
 \begin{array}{rcl}
 \hat{\epsilon_1}\frac{d\phi}{d\epsilon_1}&<&0\, ,  \\
 \hat{\epsilon_1}\frac{dV}{d\epsilon_1}&>&0\, ,
 \end{array}
 \right.
 \label{eq:epsConds}
 \end{equation}
 are fulfilled only for
 $\epsilon_1\in (0,1/B)$ and $\epsilon_1\in (1/B, \epsilon_1^{(upper)}]$.
 To conclude this section,
 therefore, the
 potential as function of the inflaton field
 is shown in
 Fig.~\ref{fig:GPLVph}.
 As can be observed in this figure,
 the required fine--tuning of $\epsilon_1$
 is equivalent to
 a fine--tuning on the initial value of the
 inflaton field,
 i.e., a sufficiently high value for $|\phi -\phi_0 |$ is
 required.
 In this plot, the branch where $\epsilon_1\in (0,1/B)$
 is
 denoted by $V_I$, and $V_{II}$ denotes the branch where
 $\epsilon_1\in (1/B, \epsilon_1^{(upper)}]$.

 \section{Discussion}

 Motivated by future prospects for
 measuring any possible contribution of the
 primordial tensor (gravitational wave)
 perturbations to the CMB power spectrum, we have
 investigated the class of inflationary
 models that result in scalar and tensor
 fluctuation spectra with a constant
 ratio of amplitudes on
 all scales, $r={\rm constant}$.
 This condition is satisfied if the
 spectral indices are equal. To lowest--order in the
 slow--roll approximation, such a condition of equality implies that
 the indices must also be independent of scale and this
 corresponds to the power law inflationary model.
 However, to next--to--leading order in the slow--roll
 approximation, the indices can be equal but
 may also exhibit a non--trivial dependence on comoving wavenumber.
 Under a self--consistent approximation, we have
 determined the functional form of this dependence
 and reconstructed parametric solutions for the inflaton potentials
 that produce such spectra.

 We
 find that there are two different
 possible potentials. Surprisingly,
 for the specific ansatz we consider,
 the power law inflationary model
 is a {\em past}, rather than a future,
 attractor, in the sense that
 at high energies (early times)
 both potentials converge  to the exponential model, but
 move away from this special case
 at low energies.
 This provides a counter example to the
 generic, lowest--order
 analysis of
 Hoffman and Turner \cite{HT} who find that power law inflation
 is a future attractor for the inflationary kinematics.
 This difference arises because the
 approach of Ref.~\cite{HT} employs a strong version
 of the slow--roll approximation to analyze the constraints
 on the inflationary evolution. In particular,
 the effects of the inflaton's
 acceleration in its equation of motion
 and its
 kinetic term in the Friedmann equation were neglected. Moreover,
 the expressions relating the spectral
 indices and the ratio of the scalar and tensor
 amplitudes to
 the potential were truncated to lowest--order. In the present work, we
 have relaxed these restrictions and employed all of the
 available
 information on the inflaton
 dynamics and truncated the
 expressions for the spectral indices to next-to-leading order. Though
 the
 result of Ref.~\cite{HT} should contain some
 of the essential features of the
 dynamics, the highly non--linear nature of the next-to-leading order
 expressions leads to important deviations from these
 lowest--order results.
 Indeed, if the strong slow-roll approximation of Ref.~\cite{HT}
 had been invoked in the present analysis,
 power law inflation would have emerged as the
 unique solution.
 However, the next-to-leading order
 dynamical analysis of
 Ref.~\cite{SLIP} indicates that in the reduced phase spaces
 for the evolution of
 $\epsilon_1$, there exists a saddle point in the
 region where $\epsilon_1$ has interesting values and
 $\Delta<0$.
 This implies that with respect to
 cosmic time (recall that $d\tau/dt<0$), attractor--like
 behavior will be characteristic only of those trajectories
 that are very close
 to the
 unstable separatrices.
 Likewise,
 the saddle point acts as a repellor for those trajectories
 that are closer to the
 stable separatrices. Inflationary dynamics
 that does not correspond to a power law
 attractor was also found for the solution
 of the second case analyzed in Ref.~\cite{generic}.

 One consequence of the behavior described in the present work
 is that a strong fine--tuning of the
 initial value of the inflaton field
 is required if
 this model is to produce spectra consistent with
 observations.
 This is different to
 what typically arises in inflationary cosmology, where
 any sensitivity to initial conditions
 is washed out by the accelerated expansion.
 If we consider Figs.~\ref{fig:GPLdDe} and \ref{fig:GPLdDk}
 once more, it can be seen
 that
 the upper branch of $\delta$ grows and becomes positive,
 thus indicating a breakdown in the next--to--leading order
 analysis. For the lower branch, $\delta$ begins to evolve
 extremely rapidly, probably indicating that
 the `running' of the spectral index, $d \Delta/ d\ln k$,
 becomes too large. Either way, observational
 constraints are difficult to satisfy.
 Thus, the potential in the region open to observation
 must be sufficiently close to the exponential (power law
 inflation) model and consequently the
 energy of the field initially stored
 in its potential must be sufficiently
 high.

 One way to satisfy this requirement
 is through the
 eternal inflation mechanism, where (large)
 quantum fluctuations in the inflaton
 can
 cause the field to diffuse up its potential \cite{eternal}.  In general,
 the condition for eternal inflation to arise is that
 \cite{eternal,generic},
 \begin{equation}
 \label{eq:Cond4EtInf}
 \frac{H^2(\epsilon_{1(i)})}{\pi m_{\rm Pl}^2 \epsilon_{1(i)}} \geq 1 \, ,
 \end{equation}
 for a given value of $\epsilon_1(t_i)$.
 Recalling that $d\tau \equiv \ d \ln H^2$, we can rewrite
 this condition as
 \begin{equation}
 \label{eq:Cond4EtInf1}
 \tau-\tau_0
 \geq
 \ln\left( \pi\epsilon_{1(i)}m_{\rm Pl}^2 \right)
 \end{equation}
 and consequently, $\tau -\tau_0$ must be
 sufficiently large
 for eternal inflation to proceed. On the other hand, it
 is clear from Fig.~\ref{fig:GPLet1} that this is equivalent
 to requiring that the initial value of $\epsilon_1$ be
 close enough to $1/B$. In this case, however, the model would
 effectively be indistinguishable from that of the power law model.
 Moreover, it could be argued that this
 restriction on $\epsilon_1 (t_i)$ itself
 represents a fine--tuning. Nevertheless,
 the attractive feature of eternal inflation is that
 only a small region of the universe need
 satisfy
 Eq. (\ref{eq:Cond4EtInf1}) for the process of
 self--reproduction to start and continue indefinitely.
 If, as is commonly assumed,
 $\epsilon_1 (t_i)$ is randomly distributed, then at any given time
 in an inflationary universe described by this model, there is a finite
 probability for the existence of
 a region satisfying the necessary condition.

 In conclusion,
 therefore,
 our analysis indicates that it is difficult, from
 a theoretical point of view, to
 obtain spectra where $r$ is truly constant
 if the spectral indices have a non--trivial scale--dependence.
 This particularly applies in the case where the slow-roll condition is
 not necessarily satisfied.
 Moreover, the potential must be
 close to the exponential form over the range of
 inflaton
 values accessible to observations.
 This implies that the observations would not
 be able to discriminate between the models
 discussed above and power law inflation.
 Consequently,
 if a running of the scalar spectral index
 is eventually favored by future observations,
 it is likely that
 the underlying theory would be much more complicated
 and, in particular, would also result in a
 non--trivial scale--dependence for the ratio
 of amplitudes.

 \acknowledgments

 C.A.T-E. and A.A.G. are supported in part by
 the CONACyT grant 38495--E and the
 Sistema Nacional de Investigadores (SNI).
 J.E.L. is supported by the Royal Society.


\begin{references}

 \bibitem{CMBdata}
 C.~B.~Netterfield {\em et al}., {\tt astro-ph/0104460};
 A.~T.~Lee {\em et al}., {\tt astro-ph/0104459};
 R.~Stompor {\em et al}., {\tt astro-ph/0105062};
 N.~W.~Halverson {\em et al}., {\tt astro-ph/0104489};
 C.~Pryke {\em et al}., {\tt astro-ph/0104490}.

 \bibitem{A_T}
 M.~Tegmark,
 Astrophys.\ J.\  {\bf 514}, L69 (1999);
 M.~Tegmark, M.~Zaldarriaga, and A.~J.~Hamilton,
 Phys.\ Rev.\ D {\bf 63}, 043007 (2001);
 X.~ Wang, M.~Tegmark, and M.~Zaldarriaga, {\tt astro-ph/0105091}.

 \bibitem{TDegen}
 G.~Efstathiou and J.~R.~Bond, Mon. Not. Roy. Astron. Soc. {\bf 304}, 75
 (1999);
 G.~Efstathiou, {\tt astro-ph/0109151};
 G.~Efstathiou {\it et al.}, {\tt astro-ph/0109152}.

 \bibitem{hudo}
 W.~Hu and S.~Dodelson, {\tt astro-ph/0110414}.

 \bibitem{SIA_O}
 B.~P.~Schmidt {\it et al.},
 Astrophys.\ J.\  {\bf 507}, 46 (1998);
 A.~G.~Riess {\it et al.}  [Supernova Search Team Collaboration],
 Astron.\ J.\  {\bf 116}, 1009 (1998).


 \bibitem{Perlmutter:1999np}
 S.~Perlmutter {\it et al.}  [Supernova Cosmology Project Collaboration],
 Astrophys.\ J.\  {\bf 517}, 565 (1999);
 S. Jha {\it et al.}, {\tt astro-ph/0101521};
 A. H. Jaffe {\em et al.}, Phys. Rev. Lett. {\bf 86}, 3475 (2001);
 R. Bean and A. Melchiorri, {\tt astro-ph/0110472}.


 \bibitem{inflation}
 A.\ Linde, {\em Particle Physics and Inflationary
 Cosmology} (Harwood, Chur, Switzerland, 1990);
 A.~R.~Liddle and D.~H.~Lyth,
 {\em Cosmological Inflation and Large-Scale Structure}
 (Cambridge University Press, Cambridge, 2000).

 \bibitem{starwave}
 A.~A.~Starobinsky, Pis'ma Zh. Eksp. Teor. Fiz.
 {\bf 30}, 719 (1979) [JETP Lett. {\bf 30}, 682 (1979)].

 \bibitem{perturbations}
 V.~Mukhanov and G.~Chibisov, Pis'ma Zh. Eksp. Teor. Fiz. {\bf 33},
 549 (1981) [JETP Lett. {\bf 33}, 532 (1981)];
 A.~H.\ ~Guth and S.~Y.~Pi, Phys. Rev. Lett. {\bf 49}, 1110 (1982);
 S.~Hawking, Phys. Lett. {\bf 115B}, 295 (1982);
 A.~A.\ ~Starobinsky, Phys. Lett. {\bf 117B}, 175 (1982);
 A.~D.\ ~Linde,
 Phys.\ Lett.\ B {\bf 116}, 335 (1982);
 J.~M.~Bardeen, P.~J.~Steinhardt, and M.~S.~Turner,
 Phys.\ Rev.\ D {\bf 28}, 679 (1983).

 \bibitem{abbott}
 L.~F.~Abbott and M.~B.~Wise,
 Nucl.\ Phys.\ B {\bf 244}, 541 (1984).

 \bibitem{TDegen1}
 M.~Zaldarriaga, D.~N.~Spergel, and U.~Seljak,
 Astrophys.\ J.\  {\bf 488}, 1 (1997).

 \bibitem{KK}
 M.\ Kamionkowsky and A.\ Kosowsky, Ann.\ Rev.\ Nucl.\ Part.
 \ Sci.\ {\bf 49}, 77 (1999).

 \bibitem{PLinfl}  F.\ Lucchin and S.\ Matarrese, Phys.\ Rev.\ D {\bf
 32}, 1316
 (1985).

 \bibitem{Lea}
 E.~J.~Copeland,  E.~W.~Kolb,
 A.~R.~Liddle, and J.~E.~Lidsey,
 Phys.\ Rev.\ D {\bf 49}, 1840 (1994);
 J.~E.~Lidsey, A.~R.~Liddle, E.~W.~Kolb, E.~J.~Copeland, T.~Barreiro, and
 M.~Abney,
 Rev.\ Mod.\ Phys.\  {\bf 69}, 373 (1997).

 \bibitem{SLIP} E.\ Ay\'on-Beato, A.\ Garc\'\i a, R.\ Mansilla, and
 C.~A.\ Terrero-Escalante, Phys.\ Rev.\ D {\bf 62}, 103513 (2000).

 \bibitem{HT}  M.~B.~Hoffman and M.~S.~Turner, Phys.\ Rev.\ D {\bf 64},
 023506
 (2001).

 \bibitem{eternal}
 P.~J.~Steinhardt, in {\em The Very Early Universe}, eds: G.~W.~Gibbons,
 S.~W.~Hawking, and S.~T.~Siklos (Cambridge University Press, Cambridge,
 1983);
 A.~Vilenkin, Phys. Rev. D {\bf 27}, 2848 (1983);
 A.~D.~Linde, Mod. Phys. Lett. A {\bf 1}, 81 (1986);
 A.~D.~Linde, Phys. Lett. {\bf 175B}, 395 (1986).

 \bibitem{HFFampl}
 D.~J.~Schwarz, C.~A.~Terrero-Escalante, and A.~A.~Garcia,
 Phys.\ Lett.\ B {\bf 517}, 243 (2001).

 \bibitem{ConstNt}
 C.~A.~Terrero-Escalante, E.~Ay\'on-Beato, and
 A.~A.~Garc\'{\i}a, Phys. Rev. D {\bf 64}, 023503 (2001).

 \bibitem{LW}
 R.~M.~Corless, G.~H.~Gonnet, D.~E.~G.~ Hare, D.~J.~Jeffrey, and
 D.~E.~Knuth,
 Adv. Comput. Math. {\bf 5}, 329 (1996).

 \bibitem{SLIPk} E.\ Ay\'on-Beato, A.\ Garc\'\i a, R.\ Mansilla, and
 C.~A.\ Terrero-Escalante, in {\em Proceedings of III DGFM-SMF Workshop
 on
 Gravitation and Mathematical Physics}, Le\'on, M\'exico (2000),
 eds: N.\ Bret\'on, O.\ Pimentel, and J. Socorro, {\tt astro-ph/0009358}.

 \bibitem{Ei}
 M.~Abramowitz and I.~Stegun,
 {\em Handbook of Mathematical Functions},
 (Dover Publications Inc., New York, 1965).

 \bibitem{generic}
 C.~A.~Terrero-Escalante and A.~A.~Garcia,
 Phys.\ Rev.\ D {\bf 65}, 023515 (2002)



 \end{references}
 \end{document}